\documentclass[12pt]{article}
\usepackage{scicite}
\usepackage{times}
\usepackage{amsmath}
\usepackage{graphicx}
\usepackage{amssymb}
\usepackage{accents}
\DeclareMathOperator{\atantwo}{atan2}
\newenvironment{sciabstract}{%
\begin{quote} \bf}
{\end{quote}}

\newcounter{lastnote}

\title{Analog Counterdiabatic Quantum Computing} 

\author
{Qi Zhang$^1$ $^{\ast}$, Narendra N. Hegade$^1$ $^{\dagger}$, Alejandro Gomez Cadavid $^{1,2}$, \\ Lucas Lassablière $^4$, Jan Trautmann$^1$, S\'{e}bastien Perseguers$^{1,3}$,\\ Enrique Solano$^1$ $^{\ddagger}$,  Loïc Henriet $^4$, Eric Michon$^1$ $^\S$\\
\\
\normalsize{$^1$ Kipu Quantum GmbH, Greifswalderstraße 226, 10405 Berlin, Germany}\\
\normalsize{$^2$ Department of Physics, University of the Basque Country UPV/EHU}\\
\normalsize{$^3$ Gradiom Sàrl, Avenue de Tivoli 4, 1700 Fribourg, Switzerland}\\
\normalsize{$^4$ PASQAL, 7 rue Léonard de Vinci, 91300 Massy, France}\\
\\
\normalsize{E-mail:  $^\ast$qi.zhang,$^\dagger$narendra.hegade,$^\ddagger$eric.michon@kipu-quantum.com}
}
\date{}

\begin{document} 

\baselineskip24pt

\maketitle 

\begin{sciabstract}
  We propose analog counterdiabatic quantum computing (ACQC) to tackle combinatorial optimization problems on neutral-atom quantum processors. While these devices allow for the use of hundreds of qubits, adiabatic quantum computing struggles with non-adiabatic errors, which are inevitable due to the hardware's restricted coherence time. We design counterdiabatic protocols to circumvent those limitations via ACQC on analog quantum devices with ground-Rydberg qubits. To demonstrate the effectiveness of our paradigm, we experimentally apply it to the maximum independent set (MIS) problem with up to 100 qubits and show an enhancement in the approximation ratio with a short evolution time. We believe ACQC establishes a path toward quantum advantage for a variety of industry use cases.
\end{sciabstract}


\section*{Introduction}

Solving combinatorial optimization problems is highly relevant in science, technology, and industry \cite{fu1986application, wurtz2022industry}, however, often these problems are computationally hard on classical computers \cite{lenstra1979computational, colorni1996heuristics}. With the recent developments in quantum computing hardware, speeding up the computation of industry-relevant problems on quantum computers is in reach \cite{MISQUERA, king2023quantum}. Typically, the Hamiltonian of a quantum system is used to encode the problem's cost function~\cite{lucas14}. A promising approach to solve the problem is to use an analog quantum computing device, where an initial quantum state is evolved into the ground state of the problem Hamiltonian via an adiabatic evolution~\cite{farhi00, albash2018adiabatic}.

Recently, arrays of neutral atoms trapped in optical tweezers have emerged into a promising hardware platform for analog quantum computing~\cite{browaeys20, scholl21, ebadi21}, besides the already available quantum annealing hardware~\cite{brooke99, king2023quantum}. The neutral-atom analog quantum computers can use hundreds of atoms where each atom serves as a qubit. The strongly interacting atomic Rydberg state~\cite{morgado21} allows the generation of entanglement, which is the heart of quantum computation. The atomic array can be configured so that the quantum many-body ground state natively encodes the solution of the maximum independent set (MIS) problem~\cite{MISQUERA, coelho22, wurtz22}, obtained via adiabatic evolution. This approach can be used to tackle industrially relevant optimization problems \cite{wurtz2022industry}. A recent proposal demonstrates how to solve non-native combinatorial optimization problems on this hardware \cite{wurtz2024solving}. However, in this finite-time adiabatic evolution, non-adiabatic errors are not avoidable due to the limited coherence time of the hardware. The errors result in reduced computation fidelity. One way to address this challenge is by finding optimal scheduling functions to describe the adiabatic evolution \cite{finvzgar2024designing}, though this can be resource-demanding and would require multiple iterations on the hardware. An alternative way to circumvent the non-adiabatic excitations is by counterdiabatic (CD) protocol as introduced in \cite{demirplak2003adiabatic, Berry2009, ChenPRL2010, del2013shortcuts}. The main idea behind CD protocols is to introduce an additional term to the fast-evolving adiabatic Hamiltonian to suppress the transition between eigenstates. However, the application of these initial proposals for CD protocols suffered from the difficulty in calculating the exact CD terms for large systems. Moreover, the required knowledge of instantaneous eigenstates to obtain the CD terms hindered its applications in adiabatic quantum computing (AQC). There have been several attempts to overcome this challenge \cite{saberi2014adiabatic}. Notably, a proposal for a variational CD protocol \cite{PNASCD, ClaeysFECD} represents significant progress in this direction. This approach offers a method to construct approximate CD terms variationally, without requiring knowledge of the Hamiltonian spectra. In this regard, several theoretical advancements have been made to improve this protocol \cite{takahashi2024shortcuts, nakahara2022counterdiabatic}, alongside experimental realizations on both digital and analog quantum processors \cite{hegade2021shortcuts, hegade2022digitized, chandarana2023digitized, guan2023single, cadavid2023efficient, hayasaka2023general}. Additionally, digital-analog methods have recently been proposed \cite{kumar2024digital}.

In this work, we introduce a method to enhance the performance of current analog quantum processors by applying analog counterdiabatic quantum computing (ACQC) techniques, specifically designed for direct implementation on the neutral atom quantum computing platform. Our method focuses on minimizing non-adiabatic errors through the introduction of CD terms, realized through the use of analytically calculated scheduling functions that control the amplitude, detuning, and phase of the driving laser used in neutral atom quantum computing experiments. This approach significantly improves the fidelity of the computation in comparison to standard adiabatic protocols. Recognizing the limitations of current hardware i.e. short coherence time, noise, and the lack of flexibility in the control variables, we tailor the CD protocols to accommodate these constraints. To demonstrate the effectiveness of our proposed CD protocols, we tackle an industrially relevant combinatorial optimization problem—the maximum independent set (MIS) problem—featuring up to 100 nodes across several instances and benchmark our results against conventional finite-time adiabatic quantum optimization protocols executed on actual hardware.  Additionally, we discuss the implementation of more advanced CD protocols on next-generation programmable neutral atom quantum hardware, equipped with individual addressing capabilities.

\section*{Results}
\label{secexp}
We introduce ACQC paradigm to solve the MIS problem on neutral atoms hardware with ground-Rydberg qubits \cite{lukin2001dipole, saffman2005analysis}.
We calculated the counterdiabatic potential analytically, taking into account the hardware's controllability, which includes one-body Pauli terms. This approach compensates for the non-adiabatic transitions of the driving part of the ground-Rydberg qubit system. We then show a way to directly implement the CD protocol on the neutral atom hardware through well-designed scheduling functions including the CD coefficients. 

\subsubsection*{Hardware implementation of ACQC}
The Hamiltonian describing the ground-Rydberg qubits is
\begin{equation}
\label{eqHryd}
\frac{H_{\text{Ryd}}(t)}{\hbar} = \underbrace{\frac{\Omega(t)}{2} \left[ \cos \varphi (t) \sum_{i=1}^N \sigma_i^x - \sin \varphi (t) \sum_{i=1}^N \sigma_i^y \right] - \Delta(t)\sum_{i=1}^N n_i}_{H_{\text{drive}}} + \underbrace{ \sum_{i< j} J_{i,j} n_i n_j}_{H_{\text{int}}},
\end{equation}
where $\Omega(t)$ is the Rabi frequency, $\Delta(t)$ is the detuning of the two-photon transition, $\varphi(t)$ is the phase of the laser, $n_i=\left| 1 \right\rangle_i \left\langle 1 \right|= (1-\sigma^z_i)/2$, and $J_{i,j}\propto r_{i,j}^{-6}$ is the interaction strength which is a function of the distance between two atoms $i$ and $j$.
To simplify the calculation, we assume $\hbar=1$ in this work.

Note that (piecewise) linear scheduling functions are the easy choice to control the Rydberg system while fulfilling boundary conditions: $\Omega(0)= 0$, $\Delta(0) = -\Delta_0$, $\Omega(T) = 0$, and $\Delta(T) =\Delta_0$, where $\Omega_0$ and $\Delta_0$ are the maximum detuning parameter satisfying the experimental limitation. 
However, generally they are not efficient at solving the MIS problem for a shorter evolution time. 
Therefore, in the rest of the manuscript, we use the linear control functions as a baseline AQC protocol to solve MIS problems.
Beyond that, a set of smooth scheduling functions for $\Omega(t)$, $\Delta(t)$ and $\varphi(t)$ are as follows,
\begin{eqnarray}
\Omega(t) &=& \Omega_0\ \sin^2\left(\frac{\pi}{2} \sin( \frac{\pi t}{T}) \right),\label{eq:OmegaSin}\\
\Delta(t) &=& -\Delta_0\ \cos(\frac{\pi t}{T}),
\label{eq:DeltaCos}\\
\varphi(t) &=& 0,
\label{eq:PhiZero}
\end{eqnarray}
which shows better performance on average compared with the linear protocol.
Based on the smooth AQC protocol in Eq. (\ref{eq:OmegaSin}-\ref{eq:PhiZero}), ACQC protocol is calculated and benchmarked to improve the results further.
Besides, since there is no strict constraint on $\varphi(t)$ to solve MIS, we start with a simple case $\varphi=0$, and the Rydberg Hamiltonian in Eq.~(\ref{eqHryd}) becomes
\begin{equation}
\label{eqHryd_phase0}
H_{\text{Ryd}}(t) = \frac{\Omega(t)}{2} \sum_i \sigma_i^x - \Delta(t)\sum_i n_i + \sum_{i< j} J_{i,j} n_i n_j,
\end{equation}
and the corresponding CD terms obtained analytically from Eq. (\ref{eqCDansatz}) following the Methods section is
\begin{equation}
\label{eqCD1b}
\left. H_{\text{CD}} (t)\right |_{J\rightarrow 0} = -\frac{\Omega \dot{\Delta}-\Delta \dot{\Omega}}{2 (\Omega^2+\Delta^2)} \sum_i \sigma_i^y.
\end{equation}
After adding the above CD terms into Eq.~(\ref{eqHryd_phase0}), the new control functions of Rydberg Hamiltonian in Eq.~(\ref{eqHryd}) become
\begin{eqnarray}
\widetilde{\Omega}(t) &=& \sqrt{g_1^2+g_2^2},\\
\widetilde{\Delta}(t) &=& \Delta,\\
\widetilde{\varphi}(t) &=& - \phi,
\end{eqnarray}
where $g_1(t) = \Omega$, $g_2(t) = -(\Omega \dot{\Delta}-\Delta \dot{\Omega})/(\Omega^2+\Delta^2)$,
and $\phi (t) = \atantwo(g_2, g_1)$.
These CD scheduling functions are used to tackle MIS problems.

\begin{figure}[t]
\centering
{\includegraphics[width=1\textwidth]{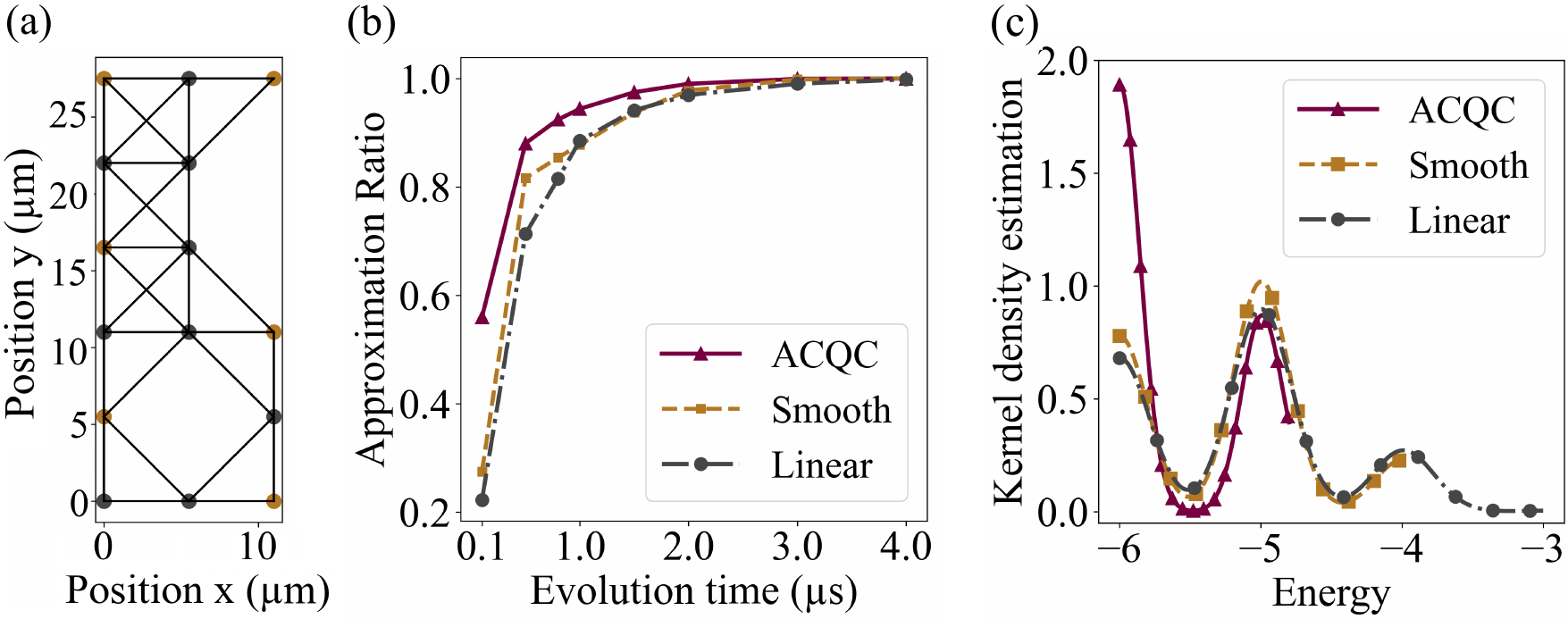}}
\caption{(a) A randomly generated graph with 15 nodes with one of the MIS solutions showing as nodes in brown. (b) The approximation ratio as a function of evolution time $T$ for solving the MIS of the graph plotted in (a) is compared between the linear protocol (linear scheduling functions, grey dash-dotted line with circles), smooth AQC (smooth scheduling functions, orange dashed line with squares) and ACQC (calculated based on the smooth AQC protocol, solid line with upper triangles). For one example of $T=1$ µs, the energy distribution analysed by using Kernel density estimation is plotted in (c). 
Parameters: $\Omega_{max}= 15$ MHz and $\Delta_{max}= 17$ MHz. The atoms are also placed on a square grid of length 5.5 µm ensuring that atoms from the same square unit and on the diagonal are within the Rydberg blockade range from each other.}
\label{fig1}
\end{figure}

\begin{figure}[t]
\centering
{\includegraphics[width=1\textwidth]{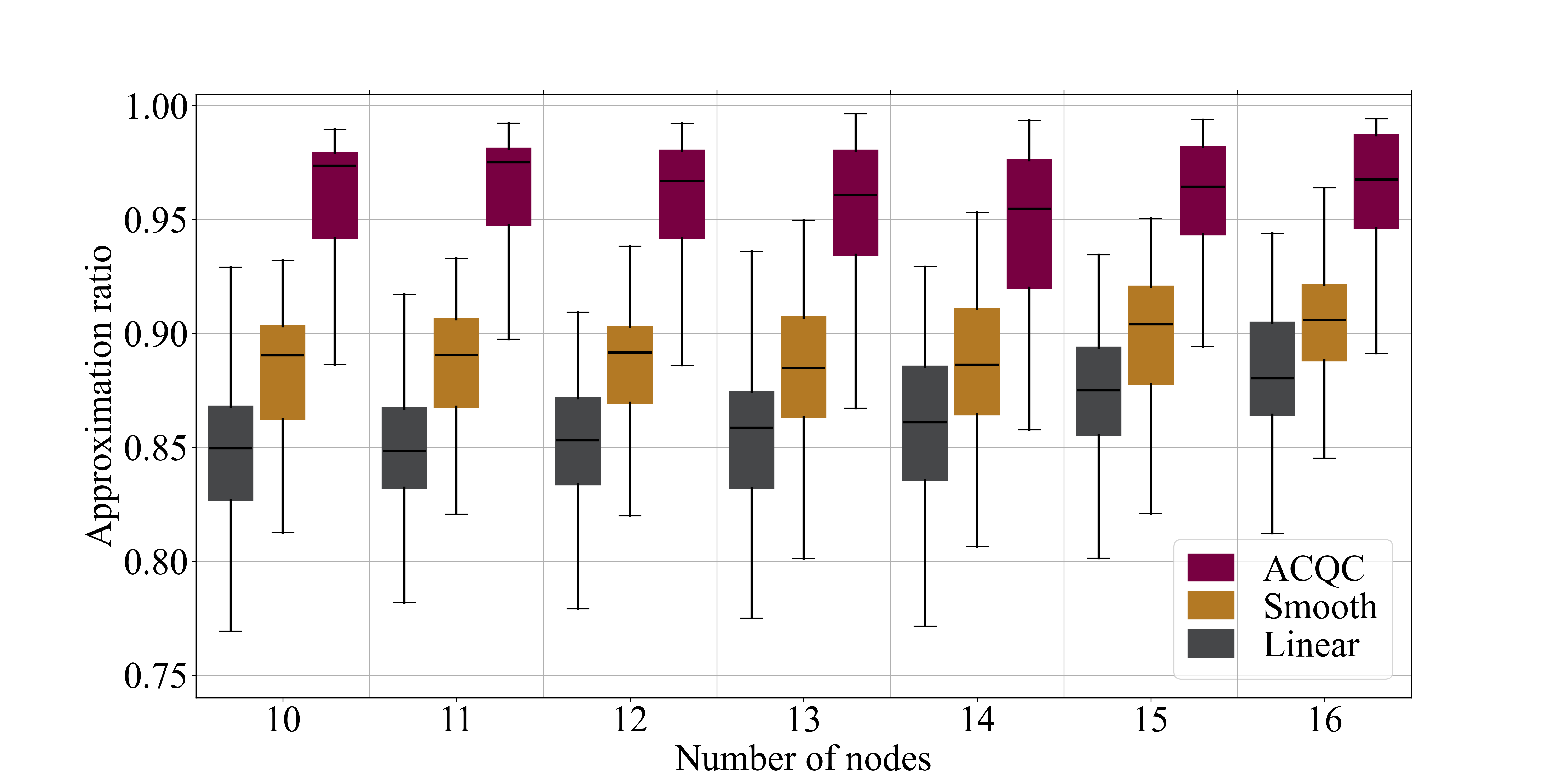}}
\caption{Noiseless simulation results for the approximation ratio for evolution time $T=1$ µs for different numbers of nodes/qubits. For each number of qubit box plot, we simulate 100 randomly generated graphs for the statistical evidence. The comparison is between the ACQC, smooth AQC and linear AQC protocols with the same parameters as in Fig.~\ref{fig1}.}
\label{fig2}
\end{figure}

\begin{figure}[h!]
\centering
{\includegraphics[width=0.9\textwidth]{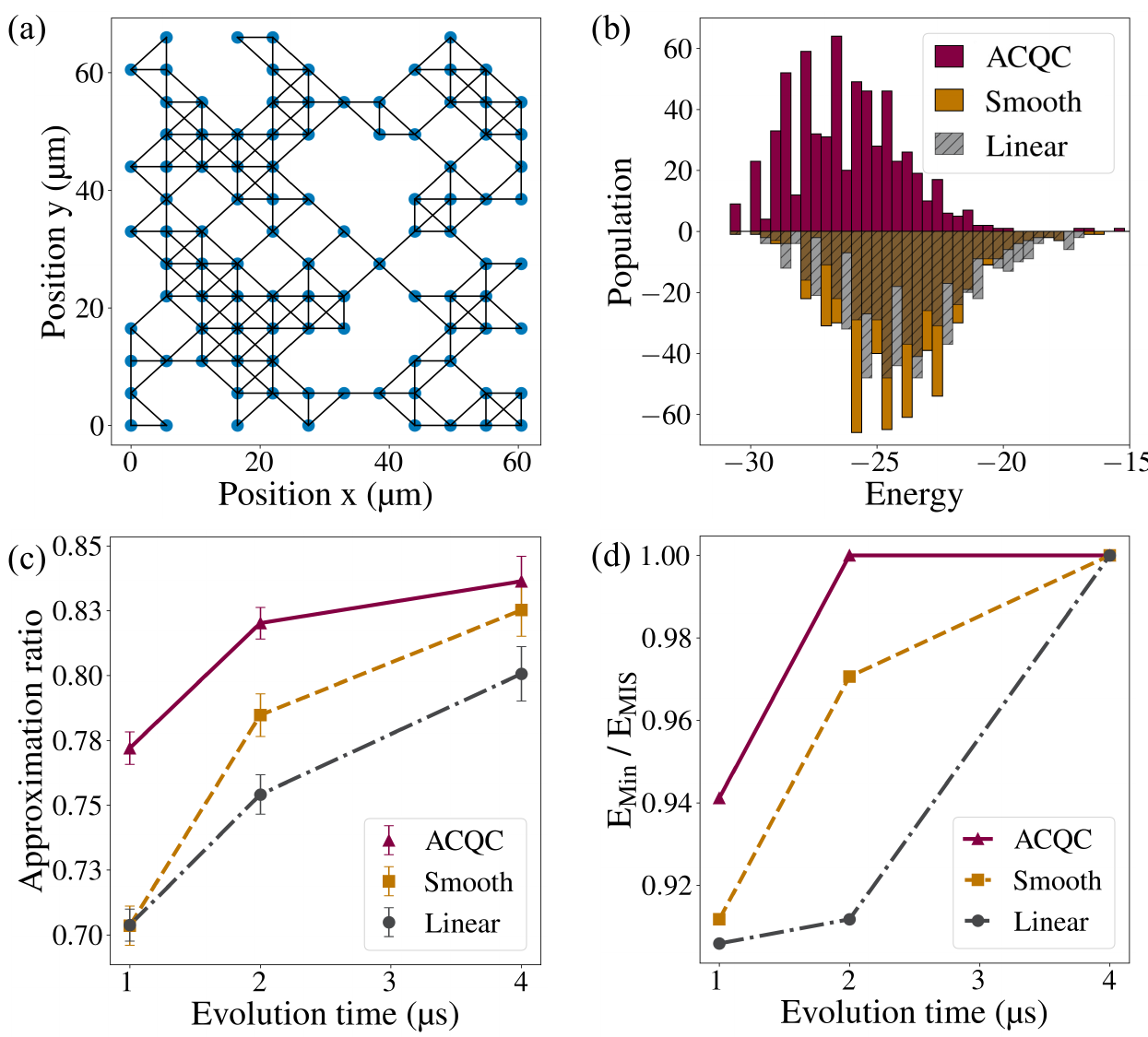}}
\caption{Experimental results obtained from solving MIS for a 100 nodes graph using QuEra's cloud-accessed Aquila quantum processor using linear scheduling functions (linear), smooth chosen scheduling functions (smooth) and our ACQC protocol calculated from the smooth AQC protocol scheduling functions. (a) Implementation of the graph onto the atomic register. The connections are drawn when two atoms are separated by less than the blockade radius from one another. (b) Distribution of bitstring energy for an evolution time of 1 µs. (c) Evolution of the approximation ratio with different evolution time, with confidence interval. (d) Evolution of the ratio between minimum energy obtained and ground state energy or the energy of the state encoding the MIS solution with different evolution time.
Parameters are the same as in Fig~\ref{fig1}.}
\label{fig3}
\end{figure}

\begin{figure}[h!]
\centering
{\includegraphics[width=0.9\textwidth]{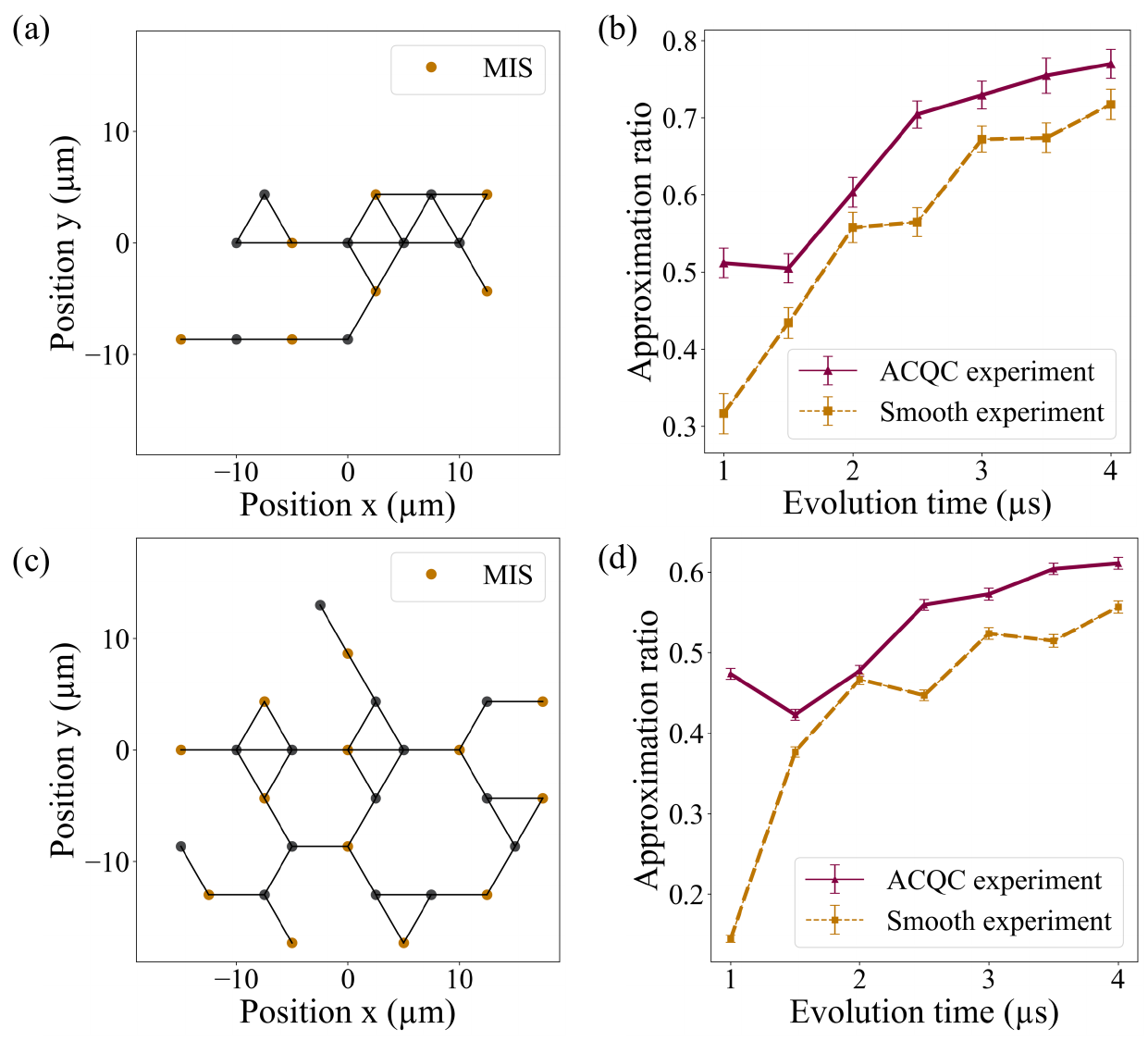}}
\caption{Experimental results obtained from solving MIS for a 15 and 27 nodes/qubits graphs, using Pasqal's Fresnel quantum processor. We use smooth schedule functions for both ACQC and the AQC protocol. (a) Mapping of the 15 nodes/qubits graph onto the atomic register. The connections are drawn when two atoms are separated by less than the blockade radius and one solution is highlighted. (b) The approximation ratio for different evolution time for the 15 nodes graph shown in (a), with confidence interval. (c) Mapping of the 27 nodes/qubits graph onto the atomic register. The connections are drawn when two atoms are separated by less than the blockade radius and one MIS solution is highlighted. (d) The approximation ratio with different evolution time for the 27 nodes/qubits graph in (c), with confidence interval.
Parameters are the same as in Fig.~\ref{fig1}.}
\label{fig3}
\end{figure}

\subsubsection*{Performance of ACQC for solving MIS problems with numerical simulations}

In order to validate the performance of ACQC, we consider MIS problems on unit-disk graphs as a use case. 
For smaller graphs, we first perform noiseless simulations.
In Fig.~\ref{fig1} (a), we show an example of a King's graph containing 15 nodes/qubits which is mapped to the atom positions on a 2-dimensional grid. We apply both AQC and ACQC protocols to solve this MIS problem, and to validate the solution, we consider approximation ratio which is the ratio between the mean energy of the output of the protocol and the lowest energy, i.e., the energy of the solution, with the following formula
$r=E_{\text{Mean}}/E_{\text{MIS}}$ as a metric where $E_\text{MIS}$ is the ground state energy encoding the MIS solution and $E_\text{Mean}$ is the mean energy of the system at the final time of the evolution. We compute the energy of each bitstring using the cost function of the MIS problem for each graph and we can then compare the mean energy of the bitstring distribution to the energy of the MIS solution to compute the approximation ratio.
In Fig.~\ref{fig1} (b), 
we plot the approximation ratio for different evolution times $T$ for both AQC and ACQC protocols.
We can see that the smooth AQC method shows better result than the linear AQC, moreover, our ACQC protocol improves on top of the smooth one.
For a shorter evolution time, taking $T=$ 0.1 µs as an example, the approximation ratio is improved from 0.222 for smooth AQC to 0.560 for ACQC with 500 shots. 
Importantly, the ACQC protocol can solve the MIS problem by reaching the ground state, however, smooth AQC has no ground state population.
For longer evolution time, the performance of AQC and ACQC start to converge which showcases the adiabatic time limit as expected.

To show statistical evidence of the superiority of ACQC over AQC, we consider 100 randomly generated graphs corresponding to MIS problems for the number of nodes between 10 and 16. The graphs are chosen to be King's graph as they are directly implementable on neutral atoms hardware. They are generated by choosing the size of the grid, the number of nodes and the probability of having a node at a given crossing. The positions are then randomly generated to create a different topology for each seed fed into the generator.
In Fig.~\ref{fig2}, we fix the evolution time $T=1$ µs and plot approximation ratio vs the increasing number of nodes/qubits for the comparison of both AQC and ACQC protocols. 
We observe an enhancement on average by 10$\%$ with ACQC.


\subsubsection*{Performance of ACQC for solving MIS problems on neutral atom hardware}

Following the clear improvement of ACQC protocol demonstrated by the simulation results, we tackle a larger MIS problem on actual hardware. We utilized the cloud-based neutral atom hardware platform provided by QuEra and Pasqal's neutral atom hardware platform as a testbed to evaluate the performance of the ACQC protocol. 
We employed the Aquila device consisting of 256 qubits, to solve an instance of the MIS problem on a 100 nodes/qubits King's graph, as shown in Fig.~\ref{fig3} (a).
The experimental results are compared across ACQC, an AQC protocol with smooth functions, and a commonly used AQC protocol with linear scheduling functions. 
Similar to the simulations, the scheduling functions for the smooth AQC are chosen 
as outlined in Eq. (\ref{eq:OmegaSin}-\ref{eq:PhiZero}), the ACQC protocol is calculated based on the same scheduling functions. Since linear scheduling function is a common choice on neutral atom hardware, it serves as a reference here. 
Having previously computed the optimal solution of the MIS problem using a classical algorithm \cite{classicalsolver, boppana1992approximating}, we can compare the optimality of the solution found through each protocol.  
We scan three different evolution times for each protocol, and each computation was performed with 1000 shots. In Fig.~\ref{fig3} (b), we compare the minimum energy obtained at different evolution time with the ground-state energy whose bitstring encodes the MIS solution. 
We also compare the mean energy in Fig.~\ref{fig3} (c) where the error bar represents the statistical confidence interval calculated from the number of shots and the standard deviation of the distribution of bitstring's energy.
As expected, for both the minimum energy and the mean energy, ACQC improves the results over AQC for a shorter time $T=1$ µs and $T=2$ µs.
For longer evolution time, $T=4$ µs, the approximation ratios for both the smooth AQC and ACQC are nearly identical and approach unity. However, the mean energy observed with ACQC is notably lower than that of the smooth AQC, and, as expected, significantly lower than the linear model.
The distribution of energy obtained for a total evolution time of 1 µs is plotted in Fig.~\ref{fig3} (d).

We perform similar experiments on Pasqal's Fresnel quantum processors, solving MIS for a 15 nodes/qubits graph and a 27 nodes/qubits graph. To showcase a different implementation of our ACQC protocol, we mapped the CD protocol using strictly the Rabi frequency and the detuning without the need for the control of the phase. 
This is done by applying a $Z-$rotation on our ACQC evolution by using a unitary operator
$U=\exp(i \varphi \sigma^z / 2)$ to control the Rydberg Hamiltonian without phase $H^{\text{Z-rot}}= (\widetilde{\Omega}^{\text{Z-rot}}(t)/2) \sum_i \sigma_i^x - \Delta^{\text{Z-rot}}(t)\sum_i n_i + \sum_{i< j} J_{i,j} n_i n_j$.
Therefore, the ACQC scheduling functions without the control of phase are $\widetilde{\Omega}^{\text{Z-rot}}(t)=\widetilde{\Omega}(t)$, 
$\widetilde{\Delta}^{\text{Z-rot}}(t)=\widetilde{\Delta}(t)+\partial_t{\widetilde{\varphi}(t)}$.
We compare performance for solving MIS using the same protocol as the previous paragraph by plottigng the approximation ratio for different evolution time, using only the smooth AQC scheduling functions and our ACQC protocol calculated from the smooth AQC functions.
For both experiments, we see the expected enhancement of ACQC. For the 27 nodes graph, at the evolution of 1 µs, the approximation ratio is improved by a factor of 3 from smooth AQC to ACQC.

 

\section*{Discussion}

We demonstrate the first implementation of a counterdiabatic protocol on a neutral atom quantum hardware for solving a combinatorial optimization problem. Our analog counterdiabatic quantum computing (ACQC) protocol is a general analytical method and is directly implementable on current commercial neutral atom hardware with ground-Rydberg qubits. 
By adding the adiabatic gauge potential of the driving part of the Rydberg Hamiltonian to the system, we can improve the results of a computation performed with a given set of scheduling functions for a time shorter than the coherence time of the system without the need for any optimization on the hardware. Not only the success probability is improved, but the mean energy of the distribution of bitstrings obtained is also lowered, making it also a useful tool for quantum sampling applications \cite{da2023quantum}.

Our ACQC protocol is readily deployable on current commercial neutral atom quantum computers, necessitating only the dynamic manipulation of Rabi frequency and detuning, alongside the dynamical adjustment of the driving laser's phase. We also propose an alternative implementation which does not require the controllability of the phase of the laser. This compatibility enables us to conduct trials via cloud access to QuEra's Aquila device as well as using Pasqal's Fresnel device, leveraging their advanced capabilities for comprehensive evaluation.

In this implementation of ACQC, the CD protocol is computed through the calculation of the adiabatic gauge potential of the Rydberg Hamiltonian in the limit of zero interactions. Because of that, at the typical maximum available evolution time of current neutral atom quantum computers, 4 µs, the ACQC protocol results starts to overlap with results obtained with an AQC protocol with well chosen scheduling functions. This is only a limitation of the particular protocol showcased in this manuscript, which constitutes an open door and a motivation for developing ansatz for CD protocols that will improve on the results presented here.

The improvement of the results through the use of ACQC for shorter computation time than 4 µs opens the door to the use of sequential processes where one part of the computation time is used to prepare a given state and the other one to perform an adiabatic computation in a time shorter than the total available time of the current hardware. ACQC would also become an important asset to perform both processes efficiently and in a compressed time.

\section*{Methods}
\subsection*{Solving combinatorial optimization problem and adiabatic quantum computing method}

Many combinatorial optimization problems, including the maximum independent set (MIS), traveling salesman problem, and quadratic assignment problem, can be efficiently encoded as a quadratic unconstrained binary optimization (QUBO) problem \cite{lucas2014ising}. The usual neutral-atom dynamics, specifically suited for tackling such optimization problems, enable the mapping of problems onto an Ising Hamiltonian. This results in a graph problem with long-range interactions among neighboring atoms, manifesting the Rydberg blockade phenomenon, preventing simultaneous excitation of adjacent atoms to the Rydberg state.

Focusing on the current neutral atom platform with ground-Rydberg qubits, this work centers on solving the MIS problem on a unit disk graph. 
Mathematically, the MIS problem is defined as finding a set \( S \) of vertices in a graph \( G = (V, E) \) such that no two vertices in \( S \) share an edge, and \( S \) is the largest set satisfying this condition. In unit disk graphs, each vertex \( v \) represents a disk of uniform radius, with an edge \( (u, v) \) existing between two vertices if and only if the corresponding disks overlap. This spatial property of unit disk graphs correlates well with the operational dynamics of neutral atom quantum processors, which can exploit their Rydberg blocked phenomena to efficiently realize this problem. The cost function corresponding to this problem is given by 
\begin{equation}
\label{eqcf}
H(x)=A \sum_{(u, v) \in E} x_u x_v-B \sum_{v \in V} x_v,
\end{equation}
where $x_v$ are binary variables, which take the value 1 if vertex $v$ is included in the independent set $S$, and 0 otherwise. The first sum penalizes any edges where both vertices are included in the set $S$ (i.e., it enforces the independence condition). The second sum rewards the inclusion of vertices in the set $S$.  $A$ and $B$ are constants where $A > B$ to ensure that the penalty for violating the independent set condition is higher than the reward for including additional vertices. The objective is to find a configuration of $x_v$ that minimizes $H(x)$. This minimization problem with qudratic terms can be mapped to finding the ground state of an Ising Hamiltonian. This can be tackled using AQC methods. 

Adiabatic quantum computing is a well-known approach for solving combinatorial optimization problems, especially when using analog quantum computing hardware. In this method, one begins by selecting an initial Hamiltonian \( H_i \), whose ground state is both known and easy to prepare. The system is then adiabatically evolved towards the problem Hamiltonian \( H_p \) by slowly changing the driving terms as defined by a time-dependent Hamiltonian \( H(t) = f(t) H_i + g(t) H_p \). For sufficiently slow evolution, the adiabatic theorem ensures that the system remains close to its ground state throughout the process.
In this case, the wave function of the system follows the instantaneous eigenstates of the Hamiltonian while the optimization solutions are encoded to be the ground state of the final Hamiltonian, which is the target state.
However, the adiabatic evolution requires long computation times which is limited by the experiment for example the coherence time of the neutral atoms system. A non-adiabatic evolution or the noise of the system can lead to excitations in the energy spectrum and can reduce the target-state fidelity, in other words, the success probability.
Therefore, we propose analog counterdiabatic quantum computing method.

\subsection*{The ground-Rydberg Hamiltonian}
Consider the Hamiltonian of neutral atoms platform using ground-Rydberg qubits in Eq.~(\ref{eqHryd}). Since the initial state of this hardware is $\left|0\right\rangle^{\otimes N}$, the initial conditions read $\Omega(0)=0$ and $\Delta(0)$ is negative.
At the final evolution time $t=T$, the Rabi frequency is back to zero and
the ground state of the final Hamiltonian $H_{\text{Ryd}}(T)$ encodes the solution of the combinatorial optimization problem, as for example minimizing the cost function in Eq.~(\ref{eqcf}). Combining the initial and target constraints, one obtains the boundary conditions of the control functions as follows
\begin{eqnarray}
\label{eqbc1}
\Omega(0) &=& 0, ~ \Delta(0) = -\Delta_1,\\
\label{eqbc2}
\Omega(T) &=& 0, ~ \Delta(T) = \Delta_2,
\end{eqnarray}
where $-\Delta_1$ and $\Delta_2$ are the minimal (negative) and maximal (positive) experimental limitations of the detuning.
Analytically, the minimal eigenvalue of the initial Hamiltonian is $0$ and the second minimal eigenvalue is $\Delta_1 > 0$.
Therefore, choosing a larger value of $\Delta_1$ can enlarge the gap between the initial ground state and the initial first excited state.
For the final constraint, the ground state encoding the solution of the MIS problems means that 
the minimal eigenvalue of $H_{\text{Ryd}} (T)$ should be $-\text{max}(N_{\text{Independent-Vertices}}) \Delta_2$
which is not fixed and depends on the specific graph and its structure.
For certain graphs, the minimal eigenvalue could be 
$-N_{\text{edges}} \Delta_2$ and the second lowest eigenvalue should be $-N_{\text{vertices}}\ \Delta_2 + \sum_{(i,j)\in \text{edges}} J_{i,j}$, which provide the following condition:
$\left(N_{\text{vertices}}-N_{\text{edges}}\right) \Delta_2 < \sum_{(i,j)\in \text{edges}} J_{i,j}$.
So in the case of $N_{\text{vertices}}>N_{\text{edges}}$, a lower value of $\Delta_2$ can enlarge the gap between the ground state and the first excited state, which can improve the success probability by using AQC protocol.

Note that no boundary condition applies to $\varphi$ since its effect vanishes at initial and final evolution times due to $\Omega$. In all standard protocols, the choice is to choose a constant phase: $\varphi\equiv0$. 

\subsection*{ACQC protocol}
\label{sec:ACQC}

Hereafter, we show a way to design the scheduling functions $\Omega(t)$, $\Delta(t)$ and $\varphi(t)$ to not only fulfill the boundary conditions, but also improve the success probability for a shorter computational time.

The idea of counterdiabaticity \cite{Berry2009, ChenPRL2010} is to add an auxiliary Hamiltonian $H_{\text{CD}}$ to the adiabatic Hamiltonian. This helps to guide the system more reliably to the desired state by preventing non-adiabatic transitions. Therefore, the Hamiltonian becomes

\begin{equation}
\label{eqHtot}
H_{\text{tot}}(t)= H_{\text{ad}}(t) + H_{\text{CD}}(t).
\end{equation}

There are different ways to add CD terms. Considering a current ground-Rydberg quantum computing platform, the Ising mode interactions exist when two neighboring atoms $i$ and $j$ are in the Rydberg states, $n_i$ and $n_j$ terms in Eq.~(\ref{eqHryd}).
In the case of many-body systems, the exact adiabatic gauge potential of the dynamic system cannot be found or the energy spectrum is too expensive to calculate, obviously a nested commutator CD terms protocol \cite{PNASCD} could be a possible solution which is a variational method to search for an approximation of adiabatic gauge potential.
However, it requires additional many-body terms which is currently not available to be added to analog neutral atoms quantum computing hardware.

To find a way around, we develop an ACQC method which does not require additional many-body interaction terms added to the quantum computing system. This can be directly implemented on the current neutral atom quantum processors without optimization or post-processing on hardware.

To ensure that the system follows the desired adiabatic path and reaches the ground state of a Hamiltonian $H_{\text{ad}}(t)$ at the final time $t=T$, the constraint for $H_{\text{CD}}(t)$ to be the solution of the adiabatic gauge potential \cite{PNASCD} of $H_{\text{ad}}(t)$ is
\begin{equation}
\label{eqgauge}
\left[ \partial_t H_{\text{ad}} - i[H_{\text{ad}}, H_{\text{CD}}], H_{\text{ad}}\right]=0.
\end{equation} 
To avoid introducing extra many-body terms beyond $\sigma^z_i \sigma^z_j$ terms of the Rydberg system, 
an efficient solution is to search for the counterdiabaticity of the independent spins under the control field where the Hamiltonian is the driving part of Rydberg Hamiltonian in Eq.~(\ref{eqHryd}), $H_{\text{ad}}(t) = H_{\text{drive}}(t)$.
Then, the adiabatic gauge potential of $H_{\text{ad}}(t)$ in the limit of zero interactions can be easily solved by choosing the following CD ansatz
\begin{equation}
\label{eqCDansatz}
\left. H_{\text{CD}} (t)\right |_{J\rightarrow 0} = f_x(t) \sum_i \sigma_i^x + f_y(t) \sum_i \sigma_i^y + f_z(t) \sum_i n_i,
\end{equation}
where the general solution of the CD coefficients $f_{x,y,z}$ in Eq.~(\ref{eqCDansatz}) can be analytically calculated directly through Eq.~(\ref{eqgauge}) as
\begin{eqnarray}
f_x(t) &=& -\frac{\Omega \dot{\Delta}-\Delta \dot{\Omega}}{2 (\Omega^2+\Delta^2)} \sin \varphi + \frac{\Omega \Delta \dot{\varphi}}{2 (\Omega^2+\Delta^2)} \cos\varphi,\\
f_y(t) &=& -\frac{\Omega \dot{\Delta}-\Delta \dot{\Omega}}{2 (\Omega^2+\Delta^2)} \cos \varphi - \frac{\Omega \Delta \dot{\varphi}}{2 (\Omega^2+\Delta^2)} \sin\varphi,\\
f_z(t) &=& \frac{\Omega^2 \dot{\varphi}}{\Omega^2+\Delta^2}.
\end{eqnarray}
Finally, the total Hamiltonian with CD terms in Eq.~(\ref{eqHtot}) should be implemented through the Rydberg Hamiltonian $\widetilde{H}_{\text{Ryd}}(t)=H_{\text{tot}}(t)$ with the updated scheduling functions as follows:
\begin{equation}
\label{eqtildeH}
\widetilde{H}_{\text{Ryd}}(t) = \frac{\widetilde{\Omega}(t)}{2} \left[ \cos \widetilde{\varphi} (t) \sum_i \sigma_i^x - \sin \widetilde{\varphi} (t) \sum_i \sigma_i^y \right] - \widetilde{\Delta}(t) \sum_i n_i + \sum_{i< j} J_{i,j} n_i n_j.
\end{equation}
Therefore, the counterdiabatic scheduling functions are calculated as
\begin{eqnarray}
\label{eqNewCtrl1}
\widetilde{\Omega}(t) &=& \sqrt{g_1^2+g_2^2},\\
\label{eqNewCtrl2}
\widetilde{\Delta}(t) &=& \Delta- \frac{\Omega^2 \dot{\varphi}}{\Omega^2+\Delta^2},\\
\label{eqNewCtrl3}
\widetilde{\varphi}(t) &=& \varphi - \phi,
\end{eqnarray}
with
\begin{eqnarray}
\label{eqNewCtrl4}
g_1(t) &=& \Omega \left(1+  \frac{\Delta \dot{\varphi}}{\Omega^2+\Delta^2}\right),\\
\label{eqNewCtrl5}
g_2(t) &=& -\frac{\Omega \dot{\Delta}-\Delta \dot{\Omega}}{\Omega^2+\Delta^2},\\
\label{eqNewCtrl7}
\phi (t) &=& \atantwo(g_2, g_1),
\end{eqnarray}
where $\atantwo(y,x)$ returns the four-quadrant inverse tangent of $y$ and $x$.
Obviously, the scheduling functions $\Omega(t)$, $\Delta(t)$ and $\varphi(t)$ are free to be chosen with respect to the boundary conditions in Eq.~(\ref{eqbc1}, \ref{eqbc2}) and the experimental limitations.
Once the scheduling functions are set, the ACQC control protocol in Eq.~(\ref{eqNewCtrl1}-\ref{eqNewCtrl7}) are designed and implemented on commercial neutral atoms hardware.


\bibliography{scibib}

\begin{thebibliography}{10}
\bibitem{classicalsolver}
We use maximum\textunderscore independent\textunderscore set function in NetworkX to obtain an approximate MIS solution for a given graph G.
\end{thebibliography}

\bibliographystyle{Science}




\section*{Acknowledgments}
We thank Johnathan Wurtz from QuEra Computing for the great discussions and comments.

\clearpage


\end{document}